\documentclass[
reprint,
superscriptaddress,
frontmatterverbose,
aps,
prl,
]{revtex4-2}

\usepackage{amsfonts,amsmath,amssymb,amsthm}
\usepackage{graphicx}
\usepackage{bm}% bold maths

\usepackage{xcolor}
\definecolor{midnightblue}{cmyk}{1,1,0,0.1}
\definecolor{forestgreen}{cmyk}{0.76,0,0.26,0.5}

\usepackage{hyperref}
\hypersetup{
    unicode=false,          % non-Latin characters in Acrobat’s bookmarks
    pdftoolbar=true,        % show Acrobat toolbar?
    pdfmenubar=true,        % show Acrobat menu?
    pdffitwindow=false,     % window fit to page when opened
    pdfstartview={FitH},    % fits the width of the page to the window
    pdftitle={My title},    % title
    pdfauthor={Author},     % author
    pdfsubject={Subject},   % subject of the document
    pdfcreator={Creator},   % creator of the document
    pdfproducer={Producer}, % producer of the document
    pdfkeywords={keyword1} {key2} {key3}, % list of keywords
    pdfnewwindow=true,      % links in new window
    colorlinks=true,        % false: boxed links; true: colored links
    linkcolor=midnightblue, % color of internal links (change box color with linkbordercolor)
    citecolor=magenta,      % color of links to bibliography
    filecolor=midnightblue, % color of file links
    urlcolor=midnightblue,   % color of external links
}

\usepackage{ulem}
\usepackage{soul}
\usepackage{multirow}
\usepackage{makecell}
\usepackage{array}
\usepackage{threeparttable}

\begin{document}

\title{Fully First-Principles Approach in Studying Topological Magnons}

\author{Xiaoqiang Liu}
\affiliation{International Centre for Quantum Design of Functional Materials, CAS Key Laboratory of Strongly-Coupled Quantum Matter Physics, and Department of Physics, University of Science and Technology of China, Hefei, Anhui 230026, China}
\affiliation{Hefei National Laboratory, University of Science and Technology of China, Hefei 230088, China}

\author{Ji Feng}
\email[Corresponding author:~]{jfeng11@pku.edu.cn}
\affiliation{International Center for Quantum Materials, School of Physics, Peking University, Beijing 100871, China}

\author{Zhenhua Qiao}
\email[Corresponding author:~]{qiao@ustc.edu.cn}
\affiliation{International Centre for Quantum Design of Functional Materials, CAS Key Laboratory of Strongly-Coupled Quantum Matter Physics, and Department of Physics, University of Science and Technology of China, Hefei, Anhui 230026, China}
\affiliation{Hefei National Laboratory, University of Science and Technology of China, Hefei 230088, China}

\author{Qian Niu}
\affiliation{International Centre for Quantum Design of Functional Materials, CAS Key Laboratory of Strongly-Coupled Quantum Matter Physics, and Department of Physics, University of Science and Technology of China, Hefei, Anhui 230026, China}

\date{\today}% It is always \today, today,
             % but any date may be explicitly specified

\begin{abstract}
We develop a fully first-principles approach for spin dynamics based on density functional perturbation theory. We demonstrate that the magnon wavefunction can be expressed by a set of electronic wavefunctions obtained from the decomposition of magnon density profile, enabling the direct calculation of magnonic quantities including Berry curvature and Chern number. As a concrete example, we show that monolayer CrI$_3$ can host topological magnons driven by spin-orbit coupling. Our model-free approach paves the way for the comprehensive studies of magnons in real materials.
\end{abstract}

\maketitle

\textit{Introduction---.} Magnons, low-energy collective excitations in ordered magnets, have attracted extensive interests in recent years due to their potential applications in dissipationless spin transport and novel magnonic devices~\cite{Chumak2015,Han2024}. Experimental measurements like inelastic neutron scattering can provide direct access to magnon energy dispersions, and thus have achieved great success in the early studies of magnons. In the last decade, it was realized that topological physics can also occur in magnons~\cite{Zhang2013,Shindou2013,Shindou2013_2,Owerre2016,McClarty2022}, and various topological magnon systems have been proposed, including magnon Chern insulators~\cite{Chisnell2015,Chen2018,Chen2021,Zhu2021}, Dirac magnons~\cite{Pershoguba2018,Yao2018,Yuan2020}, Weyl magnons~\cite{Li2016,Mook2016}, nodal-line magnons~\cite{Scheie2022}, and second-order topological magnon insulators~\cite{Mook2021}. Topologically protected chiral edge modes, whose propagation is unidirectional and robust against backscattering, can be highly promising in the field of magnonics\cite{Wang2018,Wang2020}, and nonzero Berry curvature of topological magnon bands can also produce magnon thermal Hall effect~\cite{Onose2010,Katsura2010,Matsumoto2011,Matsumoto2011_2}. However, the precise identification of topological magnon materials is very difficult experimentally, i.e., the direct detection of topological magnonic edge modes is extremely challenging due to the weak neutron-matter interaction; and experimental verification of magnonic topology via transport measurements is also hindered by the fact that magnon thermal hall effect is not quantized~\cite{Onose2010,Katsura2010,Matsumoto2011,Matsumoto2011_2}.

So far, the studies on topological magnons rely exclusively on the atomistic spin models, with parameters fitted to inelastic neutron scattering data~\cite{Chisnell2015,Chen2018,Chen2021,Zhu2021,Yao2018,Yuan2020,Scheie2022} or calculated from density functional theory~\cite{Kvashnin2020}. However, the spin model approach in real materials is limited by the assumption of localized magnetic moments as well as the rigid spin approximation~\cite{Lin2025}, and is also strongly dependent on what kind of spin-spin interactions is perceived to be important. For example, the involvement of Dzyaloshinskii-Moriya interaction~\cite{Zhang2013,Owerre2016} or Kitaev interaction~\cite{Joshi2018} gives rise to magnon Chern bands in honeycomb lattice, while the Heisenberg exchange interaction can only lead to gapless Dirac magnons. It therefore results in different interpretations of the topological nature of magnon bands and their origin in real materials like CrI$_3$~\cite{Chen2018,Chen2021,Lee2020,Ke2021}.

Therefore, it is highly desirable to develop a model-free magnon theory for real materials based on first-principles electronic structure methods to promote the study of topological magnons. At present, the simulations of magnon spectra can be achieved by computing the generalized linear susceptibility $\bm{\chi}$, using the density functional perturbation theory (DFPT)~\cite{Savrasov1998,Buczek2009,Cao2018,Gorni2022,Liu2023} or many-body perturbation theory~\cite{Olsen2021}. Even though the obtained magnon energy dispersions agree well with the experimental measurements, the magnon wavefunction is still missing and the further investigation of magnonic topology is impossible.

In this Letter, we propose a fully first-principles approach based on DFPT for the systematic studies of magnons in real materials. Compared with the conventional simulation of magnon spectrum, our approach is much more efficient in computing the magnon energies. Most importantly, the magnon density profile is obtained and decomposed into a set of electronic wavefunctions, from which magnon Berry curvatures as well as magnon Chern numbers can be calculated. As a concrete example, we identify monolayer CrI$_3$ as a magnon Chern insulator, where spin-orbit coupling is indispensable.

\textit{Equations for spin dynamics---.} The central physical quantity in first principles spin dynamics is the generalized linear susceptibility $\bm{\chi}$, which determines the variation of density $\delta \bm{\rho}$ induced by an external electromagnetic field $\bm{F}$ according to linear response theory:
\begin{equation}
\delta \rho_{\alpha}(\bm{r},\bm{q},\omega)\!=\!\sum_{\alpha'}\!\int\! d\bm{r}'\chi_{\alpha\alpha'}(\bm{r},\bm{r}',\bm{q},\omega)F_{\alpha'}(\bm{r}',\bm{q},\omega),
\label{eq:linres}
\end{equation}
where $\delta \bm{\rho}(\bm{r},\bm{q},\omega)$ and $\bm{F}(\bm{r},\bm{q},\omega)$ are the cell-periodic parts of Bloch-type functions with momentum $\bm{q}$ and $\alpha = (0,x,y,z)$. For simplicity, Eq.~(\ref{eq:linres}) is rewritten into a compact form as following:
\begin{equation}
\delta\bm{\rho}(\bm{q},\omega)=\bm{\chi}(\bm{q},\omega)\bm{F}(\bm{q},\omega).
\end{equation}

A magnon is then an excited state with quasimomentum $\hbar\bm q$ and energy $\hbar\omega_{s\bm q}$ satisfying
\begin{equation}
\label{eq:eigen1}
\bm{\chi}^{-1}(\bm{q},\omega_{s\bm{q}})\delta\bm{\rho}(\bm{q},\omega_{s\bm{q}})=0.
\end{equation}
Here, $s$ is the branch index and the corresponding magnon density profile is $\delta\bm{\rho}(\bm{q},\omega_{s\bm{q}})$. In general, $\bm{\chi}$ is a large non-sparse matrix in first-principles calculations \footnote{could be as large as $10^6$ even though the pseudopotential techniques are employed.} and computing its inverse is clearly prohibitive. Furthermore, only one column of $\bm{\chi}$ can be acquired in a typical self-consistent DFPT calculation, which makes the computation of the whole matrix $\bm{\chi}$ very time-consuming. Thus, it is a formidable task to solve Eq.~(\ref{eq:eigen1}) directly.

In DFPT, the generalized susceptibility $\bm{\chi}$ satisfies the Dyson-like equation
\begin{equation}
\bm{\chi}^{-1}(\bm{q},\omega)=\left[\bm{\chi}^{0}\right]^{-1}(\bm{q},\omega)-\bm{f}(\bm{q}),
\end{equation}
where $\bm{f}$ is the interaction kernel under the adiabatic local density approximation~\cite{Runge1984,Gross1985} and the Kohn-Sham susceptibility $\bm{\chi}^{0}$ is the linear response function of non-interacting Kohn-Sham system. Eq.~(\ref{eq:eigen1}) can then be transformed into
\begin{equation}
\label{eq:eigen2}
\bm{\chi}^{0}(\bm{q},\omega_{s\bm{q}})\bm{f}(\bm{q})\delta\bm{\rho}(\bm{q},\omega_{s\bm{q}})=\delta\bm{\rho}(\bm{q},\omega_{s\bm{q}}),
\end{equation}
where neither matrix inversion nor $\chi$ is needed, indicating that self-consistent DFPT calculations can be bypassed entirely. The primary task of this Letter is to solve Eq.~(\ref{eq:eigen2}), which will be elaborated later.

\textit{Magnon wavefunctions---.} Magnon wavefunctions are essential in the study of magnonic topology. However, the magnon density profile $\delta\bm{\rho}(\bm{q},\omega_{s\bm{q}})$ obtained from Eq.~(\ref{eq:eigen2}) cannot be identified with the magnon wavefunction, as $\delta\bm{\rho}(\bm{q},\omega_{s\bm{q}})$ for different branches are eigenvectors of different matrices $\bm{\chi}^{0}(\bm{q},\omega_{s\bm{q}})\bm{f}(\bm{q})$ and cannot form a orthonormal basis set. Noticing that magnons are collective excitations of many-electron systems, the magnon wavefunction should be able to be expressed by a set of electronic wavefunctions that yield $\delta\bm{\rho}(\bm{q},\omega_{s\bm{q}})$. Furthermore, since magnons are composed of spin-flip electron-hole excitations, the desired electronic wavefunction $|\delta u_{n\bm{k}}(\bm{q},\omega_{s\bm{q}})\rangle$ can be expanded within the unoccupied manifold
\begin{equation}
\left|\delta u_{n\bm{k}}(\bm{q},\omega_{s\bm{q}})\right\rangle =\sum_{m\in\mathrm{unocc}}\left|u_{m\bm{k}+\bm{q}}\right\rangle \delta\rho_{nm\bm{k}}(\bm{q},\omega_{s\bm{q}}),
\end{equation}
where $\delta\rho_{nm\bm{k}}(\bm{q},\omega_{s\bm{q}})$ is the probability amplitude of electronic transition from the occupied state $\left|u_{n\bm{k}}\right\rangle$ to the unoccupied state $\left|u_{m\bm{k}+\bm{q}}\right\rangle$.

In below, we display the procedure in our approach to obtain the magnon wavefunction from the acquired $\delta\bm{\rho}(\bm{q},\omega_{s\bm{q}})$. We first define a potential $\bm{V}(\bm{q},\omega_{s\bm{q}})=\bm{f}(\bm{q})\delta{\bm{\rho}}(\bm{q},\omega_{s\bm{q}})$ and then calculate $|\delta u_{n\bm{k}}(\bm{q},\omega_{s\bm{q}})\rangle$ by solving the following Sternheimer equation:
\begin{equation}
\label{eq:dec1}
(\hbar\omega_{s\bm{q}}-H_{\bm{k}+\bm{q}}^{0}+\varepsilon_{n\bm{k}}^{0})|\delta u_{n\bm{k}}(\bm{q},\omega_{s\bm{q}})\rangle=\hat{P}_{\bm{k}+\bm{q}}\delta H(\bm{q},\omega_{s\bm{q}})|u_{n\bm{k}}\rangle,
\end{equation}
where $\delta H(\bm{q},\omega_{s\bm{q}})=\sum_{\alpha}\sigma_{\alpha}V_{\alpha}(\bm{r},\bm{q},\omega_{s\bm{q}})$ and $\hat{P}_{\bm{k}}$ is the projector on to the unoccupied manifold with wave-vector $\bm{k}$. $H_{\bm{k}+\bm{q}}^{0}$, $\varepsilon_{n\bm{k}}^{0}$ and $|u_{n\bm{k}}\rangle$ are Kohn-Sham Hamiltonian, energies and spinor wavefunctions, respectively. Based on linear response theory, it can be proved that
\begin{equation}
\label{eq:dec2}
\begin{aligned}
\delta\bm{\rho}(\bm{q},\omega_{s\bm{q}})=&\bm{\chi}^{0}(\bm{q},\omega_{s\bm{q}})\bm{V}(\bm{q},\omega_{s\bm{q}})\\
=&\sum_{n\bm{k}}f_{n\bm{k}}[ u_{n\bm{k}}^{\dagger}(\bm{r})\sigma_{\alpha}\delta u_{n\bm{k}}(\bm{r},\bm{q},\omega_{s\bm{q}})\\
+&\delta u_{n\bm{k}}^{\dagger}(\bm{r},-\bm{q},-\omega_{s\bm{q}})\sigma_{\alpha}u_{n\bm{k}}(\bm{r})].
\end{aligned}
\end{equation}

However, it is still unclear how to define the geometry of these collective magnon excitations. Here, we define $\Psi_{s\boldsymbol q} = [\delta u(\boldsymbol{q},\omega_{s\bm{q}});\delta u^*(-\boldsymbol{q},-\omega_{s\bm{q}})]$, where $\delta u(\boldsymbol{q},\omega_{s\bm{q}})$ is a column vector formed by concatenating $|\delta u_{n\boldsymbol{k}}(\boldsymbol{q},\omega_{s\bm{q}})\rangle$ over all occupied $(n,\boldsymbol{k})$. It is straightforward to show that Eq.~(\ref{eq:dec1}) is equivalent to
\begin{equation}
    \begin{bmatrix}
        \mathcal H(\boldsymbol{q}) & \Delta(\boldsymbol{q})\\
        \Delta^\dagger(\boldsymbol{q}) & \mathcal H^*(-\boldsymbol{q})
    \end{bmatrix}
    \Psi_{s\boldsymbol{q}}
    =\hbar\omega_{s\bm{q}} \sigma^3\Psi_{s\boldsymbol{q}},
\end{equation}
where $\sigma^\alpha$ are Pauli matrices corresponding to $\pm(\boldsymbol{q},\omega)$. $\mathcal H(\boldsymbol{q})$ and $\Delta(\boldsymbol{q})$ are nonlocal operator matrices
\begin{equation}
\begin{aligned}
\mathcal H_{n\bm{k},n'\bm{k}'}(\bm{q}) =&\delta_{nn'}\delta_{\bm{k}\bm{k}'}(H_{\bm{k}+\bm{q}}^{0}-\varepsilon_{n\bm{k}}^{0})+\\
&\sum_{\alpha\alpha'}\sigma_{\alpha}u_{n\bm{k}}(\bm{r})f_{\alpha\alpha'}(\bm{r},\bm{r}',\bm{q})u_{n'\bm{k}'}^{\dagger}(\bm{r}')\sigma_{\alpha'},\\
\Delta_{n\bm{k},n'\bm{k}'}(\bm{q}) =&\sum_{\alpha\alpha'}\sigma_{\alpha}u_{n\bm{k}}(\bm{r})f_{\alpha\alpha'}(\bm{r},\bm{r}',\bm{q})u_{n'\bm{k}'}^{T}(\bm{r}')\sigma_{\alpha'}^{T}.
\end{aligned}
\end{equation}
It is then clear that magnon wavefunctions at $\boldsymbol{q}$ should be orthonormalized according to
\begin{equation}
\label{eq:normal}
    \langle \Psi_{s\boldsymbol{q}}|\sigma^3|\Psi_{s'\boldsymbol{q}}\rangle =\delta_{ss'} \sigma_s,
\end{equation}
where $\sigma_s=\pm1$ for the positive/negative energy branches, respectively. Magnon Berry connection $\bm{A}$ and magnon Berry curvature $\bm{\Omega}$ can thus be expressed in terms of so-normalized magnon wavefunctions as
\begin{equation}
\begin{aligned}
\bm{A}_{s\bm{q}}=&i\sigma_s\sum_{n\bm{k}}f_{n\bm{k}}[\left\langle \delta u_{n\bm{k}}(\bm{q},\omega_{s\bm{q}})\right|\partial_{\bm{q}}\left|\delta u_{n\bm{k}}(\bm{q},\omega_{s\bm{q}})\right\rangle \\
&-\left\langle \delta u_{n\bm{k}}^{*}(-\bm{q},-\omega_{s\bm{q}})\right|\partial_{\bm{q}}\left|\delta u_{n\bm{k}}^{*}(-\bm{q},-\omega_{s\bm{q}})\right\rangle],\\
\bm{\Omega}_{s\bm{q}}=&i\sigma_s\sum_{n\bm{k}}f_{n\bm{k}}[\left\langle \partial_{\bm{q}}\delta u_{n\bm{k}}(\bm{q},\omega_{s\bm{q}})\right|\times\left|\partial_{\bm{q}}\delta u_{n\bm{k}}(\bm{q},\omega_{s\bm{q}})\right\rangle \\
&-\left\langle \partial_{\bm{q}}\delta u_{n\bm{k}}^{*}(-\bm{q},-\omega_{s\bm{q}})\right|\times\left|\partial_{\bm{q}}\delta u_{n\bm{k}}^{*}(-\bm{q},-\omega_{s\bm{q}})\right\rangle].
\end{aligned}
\end{equation}
These are the main results of this Letter. Remarkably, at this level of the theory both the transition amplitude $\delta\rho_{nm\bm{k}}(\bm{q},\omega_{s\bm{q}})$ and the unoccupied electronic state $\left|u_{m\bm{k}+\bm{q}}\right\rangle$ contribute to the magnon Berry curvature, wherein the latter is inaccessible in the widely-adopted Heisenberg-type spin models.

\begin{figure}[tb]
    \centering
    \includegraphics[width=85 mm]{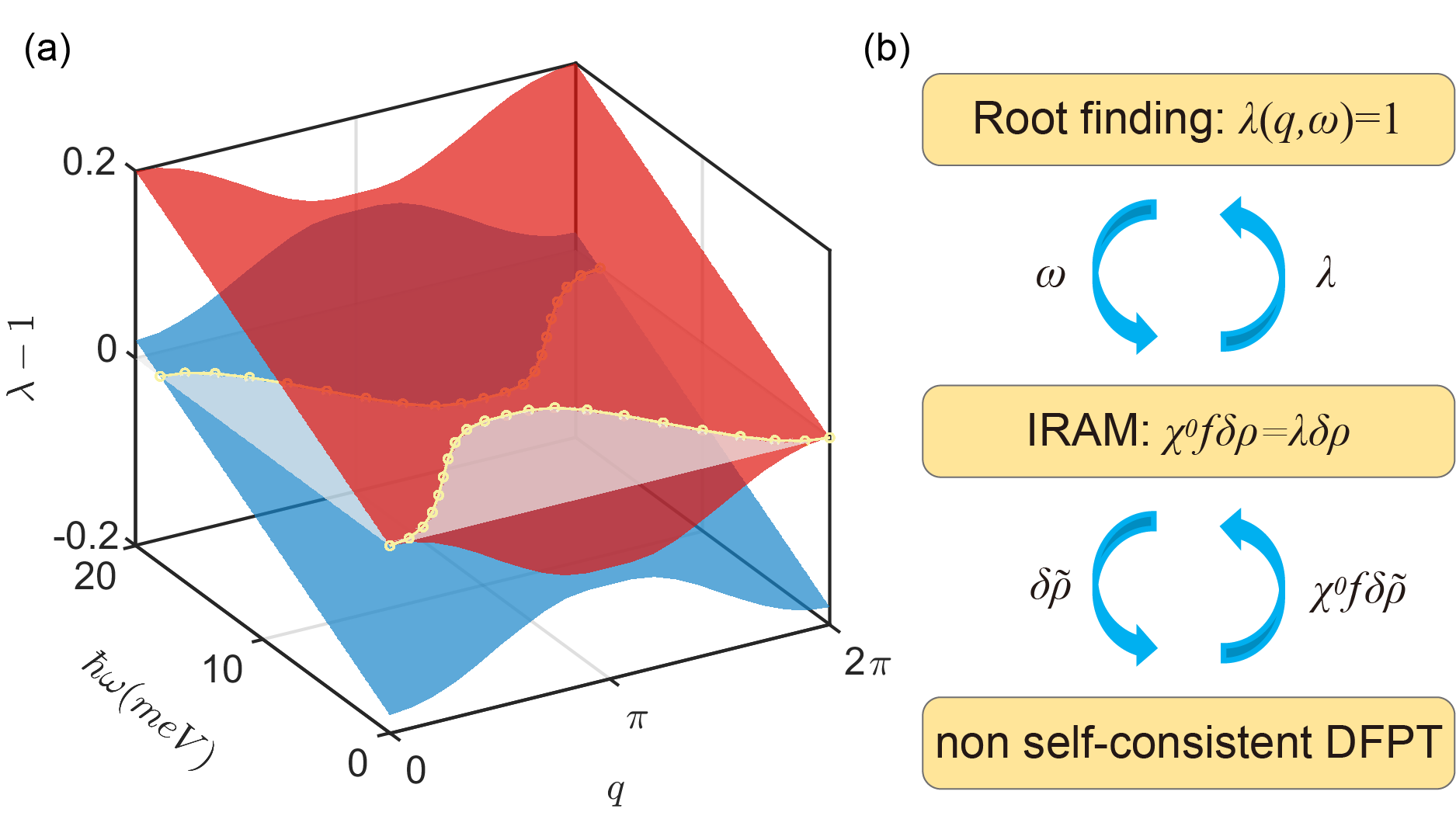}
    \caption{(a) Schematic diagram of $\lambda_{n}(\bm{q},\omega)$ in $(\bm{q},\omega)$ parameter space. The red and blue hypersurfaces are the top two $\lambda(\bm{q},\omega)$ while the white plane is $\lambda=1$ plane. The yellow intersections correspond to the energy dispersions of collective magnon excitations. (b) Computational procedure of our approach.}
    \label{fig:flow}
\end{figure}

\textit{Computational implementation---.} We present our Computational implementation for solving Eq.~(\ref{eq:eigen2}). Since $\hbar\omega_{s\bm{q}}$ is undetermined and needs to be solved at the same time, it is more convenient to introduce an auxiliary variable $\lambda$ to be the eigenvalue of matrix $\bm{\chi}^{0}\bm{f}$:
\begin{equation}
\label{eq:eigen3}
\bm{\chi}^{0}(\bm{q},\omega)\bm{f}(\bm{q})\delta\bm{\rho}_{n}(\bm{q},\omega)=\lambda_{n}(\bm{q},\omega)\delta\bm{\rho}_{n}(\bm{q},\omega).
\end{equation}
Each $\lambda_n(\bm q,\omega)$ can be seen as a hypersurface, as shown schematically in Fig.~\ref{fig:flow}(a). The desired excited states with dispersion relation $\omega=\omega_{s\bm q}$ and density profile $\delta\bm{\rho}(\bm q,\omega_{s\bm q})$ are obtained as the intersection given by $\lambda_{s}(\bm{q},\omega_{s\bm{q}})=1$. In general, only a few hypersurfaces on the top (with the largest $\lambda$) intersect with $\lambda=1$ plane. Therefore, we employ the implicitly restarted Arnoldi method (IRAM)~\cite{Sorensen1992} from the ARPACK library~\cite{Lehoucq1998} to solve Eq.~(\ref{eq:eigen3}) for the largest few eigenvalues.

A critical step in IRAM is the Arnoldi factorization that constructs an orthogonal basis for a Krylov subspace, where one needs to calculate the action of matrix $\bm{\chi}^{0}\bm{f}$ on a given vector $\delta \tilde{\bm{\rho}}$. The direct calculation and storage of matrix $\bm{\chi}^{0}$ is rather difficult in practice, and the truncation of unoccupied bands or plane-wave basis set give rise to an artificial spin excitation gap~\cite{Skovhus2021}. Instead, we calculate $\bm{\chi}^{0}\bm{f}\delta \tilde{\bm{\rho}}$ using Eqs.~(\ref{eq:dec1}) and (\ref{eq:dec2}), with the potential being replaced by $\tilde{\bm{V}}=\bm{f}\delta \tilde{\bm{\rho}}$. This is indeed a non self-consistent DFPT calculation for potential $\tilde{\bm{V}}$, which is the key to the realization of our approach and is conducted using our DFPT code~\cite{Liu2023} implemented on VASP 5.4.4 \cite{Kresse1996}. Our computational procedure is shown as a flowchart in Fig.~\ref{fig:flow}(b) and the Brent's method \cite{brent1973} is adopted to efficiently solve the root finding problem $\lambda_{s}(\bm{q},\omega_{s\bm{q}})=1$.

\textit{Topological magnons in monolayer CrI$_3$---.} We now apply our approach in monolayer CrI$_3$, which is the first material with two-dimensional ferromagnetism being observed~\cite{Huang2017}, and is predicted to be a magnon Chern insulator~\cite{Chen2018,Chen2021,Lee2020}. In our consideration, the experimental lattice parameter $a=6.867~$\AA~\cite{McGuire2015} is used, with a vacuum of about 20~\AA~to avoid the spurious interaction between layers. Planewave energy cutoff is set to $400$ eV along with a $5\times5\times1$ $\Gamma$-centered mesh for $\bm k$-points. The change of magnon energy caused by increasing the energy cutoff to $600$ eV or the $\bm k$-points mesh to $7\times7\times1$ is smaller than $0.1$ meV. Adiabatic local density approximation as well as projector augmented wave method~\cite{Blochl1994,Kresse1999} are adopted in all our calculations.

\begin{figure}[tb]
    \centering
    \includegraphics[width=85 mm]{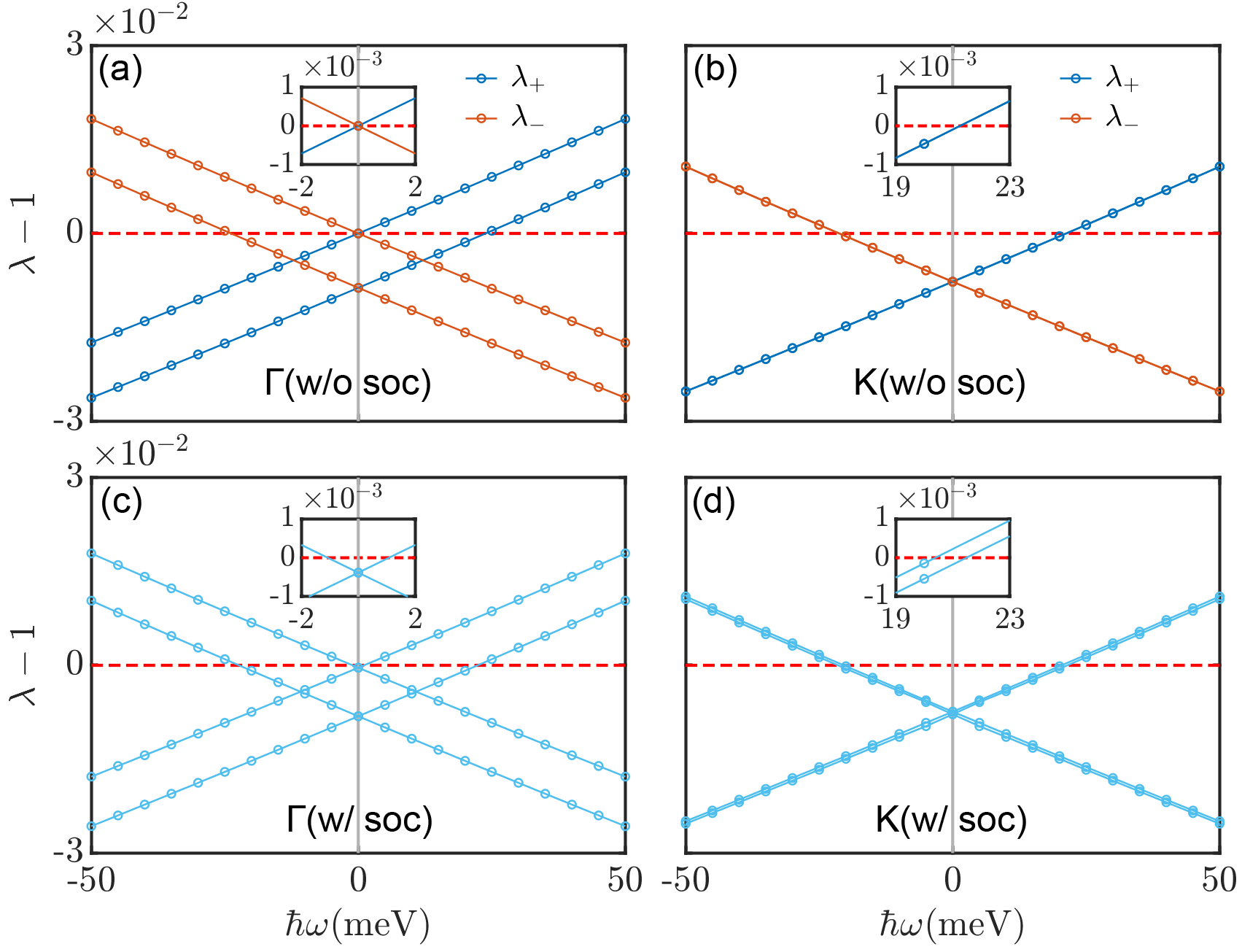}
    \caption{The calculated $\lambda(\Gamma,\omega)$ and $\lambda(K,\omega)$ in the absence (a,b) and presence (c,d) of spin-orbit coupling for monolayer CrI$_3$. The insets show the enlargement near the intersections.}
    \label{fig:lambda}
\end{figure}

Figure~\ref{fig:lambda} displays the computed $\lambda(\bm{q},\omega)$ at $\Gamma$ and $K$ point for monolayer CrI$_3$. One can see that, in the absence of spin-orbit coupling, the two transverse channels, describing the creation and annihilation of magnon respectively, are decoupled from the other remaining channels and can be calculated separately:
\begin{equation}
\begin{aligned}
\bm{\chi}_{+-}^{0}\bm{f}_{-+}\delta\bm{\rho}_{+}=&\lambda_{+}\delta\bm{\rho}_{+} \\
\bm{\chi}_{-+}^{0}\bm{f}_{+-}\delta\bm{\rho}_{-}=&\lambda_{-}\delta\bm{\rho}_{-},
\end{aligned}
\end{equation}
where $\delta\bm{\rho}_\pm =\delta\bm{\rho}_x\pm\mathrm i\delta\bm{\rho}_y$. As shown in Fig.~\ref{fig:lambda}(a) and \ref{fig:lambda}(b), $\lambda_{+}$ and $\lambda_{-}$ are symmetric with respect to $\omega=0$. We then focus on the $\lambda_{+}$ channel and it can be seen that $\lambda_{+}$ increases with increasing $\omega$. Since ground state stability implies $\lambda(\bm{q},0)\leq1$, the energy of intersection between $\lambda_{+}$ and $\lambda=1$ must be nonnegative and determines magnon energy as expected. In the low-energy region, there are two intersections for monolayer CrI$_3$, correspondingly to two magnetic atoms in a primitive cell. For $\Gamma$, one of the intersections is located exactly at 0 meV, corresponding to the Goldstone mode, while for $K$, the two eigenvalues are two-fold degenerate, indicating the existence of gapless Dirac magnons. When spin-orbit coupling is included, spin is no longer a good quantum number. Even so, most features of $\lambda$ are maintained, as shown in Figs.~\ref{fig:lambda}(c) and \ref{fig:lambda}(d). The influence of spin-orbit coupling is most prominent at $\Gamma$ and $K$:  the Goldstone mode becomes massive with a magnon excitation gap and the degeneracy of the Dirac points at $K$ is lifted.

\begin{figure}[tb]
    \centering
    \includegraphics[width=85 mm]{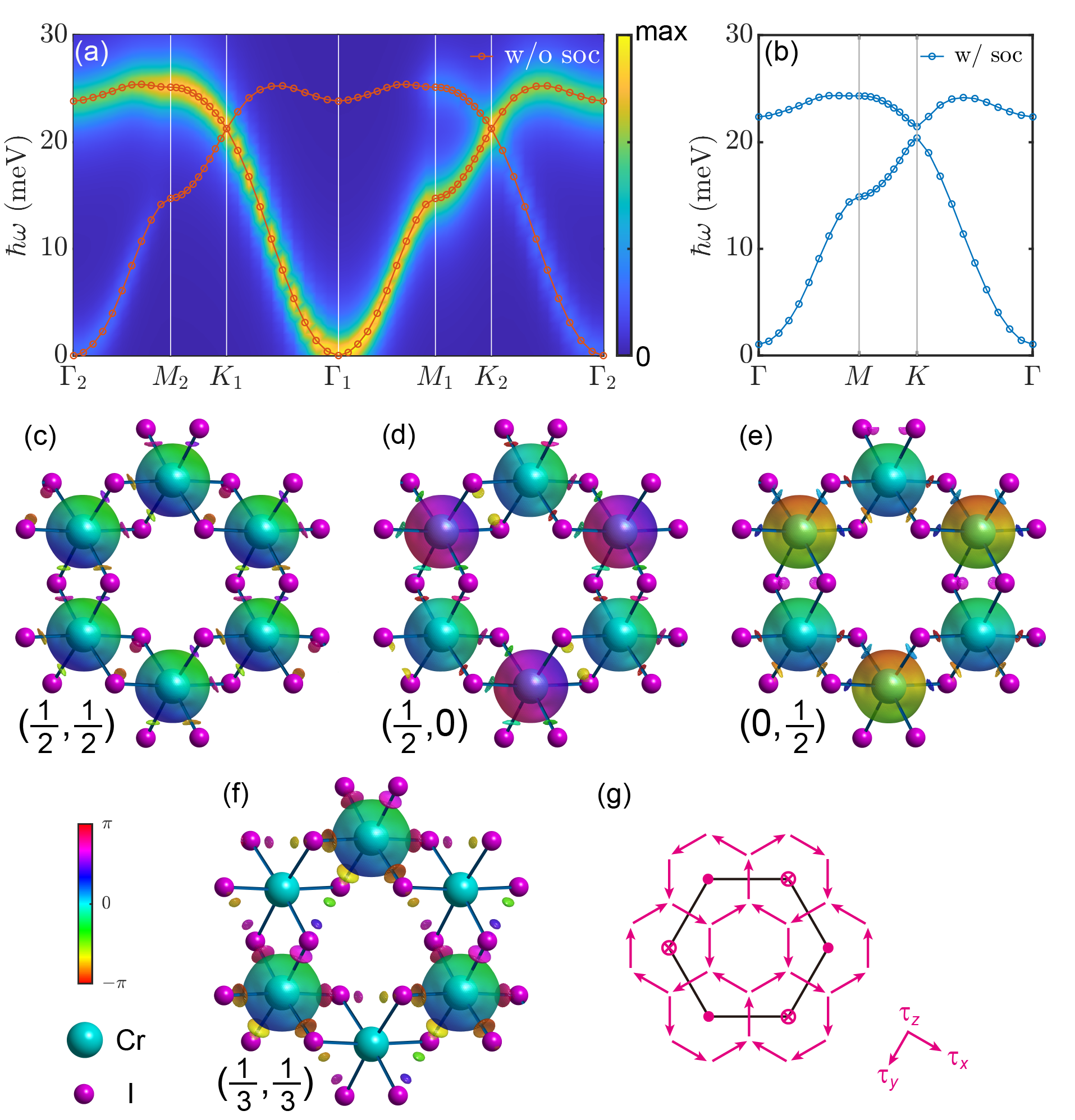}
    \caption{(a) Magnon dispersion along high symmetry path for monolayer CrI$_3$ in the absence of spin-orbit coupling. Orange line is calculated using our approach and the magnon spectrum $\mathrm{Im} \bm{\chi}(\bm{q},\omega)$ is obtained from the fully self-consistent DFPT calculations. Here $\Gamma_1=(0,0)$, $\Gamma_2=(0,1)$, $M_1=(-\frac{1}{2},\frac{1}{2})$, $M_2=(\frac{1}{2},\frac{1}{2})$, $K_1=(\frac{1}{3},\frac{1}{3})$ and $K_2=(-\frac{1}{3},\frac{2}{3})$. (b) Magnon dispersion for monolayer CrI$_3$ in the presence of SOC. (c-f) Distributions of the spin-raising component $|\delta\bm{\rho}_{+}(\bm{q},\omega_{s\bm{q}})|$ around and at $K$ for the upper band in the presence of spin-orbit coupling. The isosurfaces are colored according to the phase of $\delta\bm{\rho}_{+}(\bm{q},\omega_{s\bm{q}})$ with the average phase on one Cr atom fixed to zero. (g) Pseudospin texture for the upper band in monolayer CrI$_3$.}
    \label{fig:wq}
\end{figure}

Figures~\ref{fig:wq}(a) and \ref{fig:wq}(b) display the first-principles magnon dispersion of monolayer CrI$_3$. In the absence of spin-orbit coupling, the Goldstone mode at $\Gamma$ and the Dirac magnons at $K$ are clearly observed in Fig.~\ref{fig:wq}(a). In the presence of spin-orbit coupling, band gaps about 1.06 meV at $\Gamma$ and 1.09 meV at $K$ are respectively opened as displayed in Fig.~\ref{fig:wq}(b). As a benchmark, we also perform fully self-consistent DFPT calculations to compute $\bm{\chi}(\bm{q},\omega)$, with the magnon spectrum $\mathrm{Im} \bm{\chi}(\bm{q},\omega)$ being displayed in Fig.~\ref{fig:wq}(a). One can see that the obtained magnon dispersions from both methods are completely consistent. However, it is noteworthy that our proposed approach is much more efficient than the simulation of magnon spectrum. One can also see that the optical magnons near $\Gamma_1$ is unobservable in the spectrum and thus the calculations of $\bm{\chi}(\bm{q},\omega)$ at higher order Brillouin zones are inevitable. In addition, high-resolution magnon spectrum requires $\bm{\chi}(\bm{q},\omega)$ with a small energy interval (i.e., 1 meV in this work). In contrast, our approach only needs to be conducted at the first Brillouin zone and computing $\lambda(\bm{q},\omega)$ for a few energies (5-6 typically) is usually enough to achieve a tolerance of 0.1 meV. Even for a given pair of $(\bm{q},\omega)$, time consumption for the computation of $\bm{\chi}(\bm{q},\omega)$ is 2-5 times greater than that of $\lambda(\bm{q},\omega)$, depending on the difference between $\hbar\omega$ and magnon energy.

Although we have shown that spin-orbit coupling indeed opens a gap at $K$ in monolayer CrI$_3$, topological properties of the magnon bands remain unknown. To gain preliminary insights into the magnonic topology, we visualize the calculated magnon density profiles around and at $K$ for the upper band in Figs.~\ref{fig:wq}(c)-~\ref{fig:wq}(f), which are impossible in the conventional simulation of magnon spectrum. Here, we only display the spin-raising component $\delta\bm{\rho}_{+}(\bm{q},\omega_{s\bm{q}})$, since the magnitudes of other components are quite small. For the several selected $\bm{q}$ around $K$, $\delta\bm{\rho}_{+}(\bm{q},\omega_{s\bm{q}})$ is distributed almost equally around the two Cr atoms, with the phase difference varying; while at $K$, $\delta\bm{\rho}_{+}(K,\omega_{s\bm{q}})$ is mostly localized at one Cr atom. Pseudospin can be defined as $\bm{\tau}_{s\bm{q}}=\langle\phi_{s\bm{q}}|\bm{\sigma}|\phi_{s\bm{q}}\rangle$, where $\phi^{A/B}_{s\bm{q}}$ is the integral magnon density on A/B sublattice
\begin{equation}
\phi^{A/B}_{s\bm{q}}= \int_{\Omega_{A/B}} d\bm{r} \delta\rho_{+}(\bm{r},\bm{q},\omega_{s\bm{q}}).
\end{equation}
The pseudospin texture for the upper band is shown in Fig.~\ref{fig:wq}(g) and vortex-like structures at $K$ and $K'$ are clearly seen.

\begin{figure}[tb]
    \centering
    \includegraphics[width=85 mm]{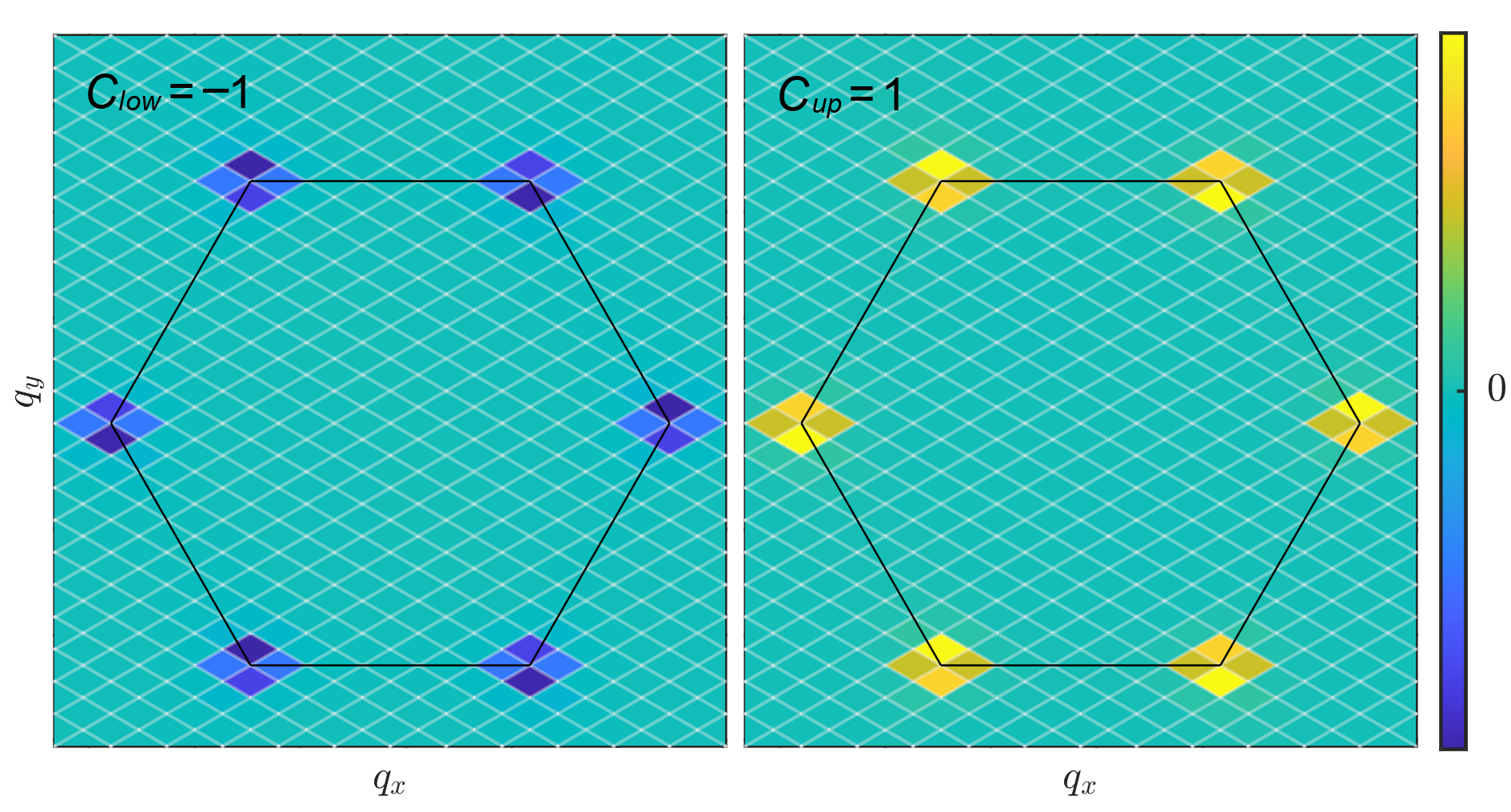}
    \caption{Distributions of Berry curvatures for the two magnon bands of monolayer CrI$_3$ in the presence of spin-orbit coupling. The black lines indicate the first Brillouin Zone.}
    \label{fig:berry}
\end{figure}

We proceed to calculate the magnon Berry curvatures and subsequently the magnon Chern numbers for the two magnon bands of monolayer CrI$_3$ in the presence of spin-orbit coupling. The magnon wavefunction is obtained by decomposing $\delta\bm{\rho}(\bm{q},\omega_{s\bm{q}})$ using the Sternheimer equation and normalized according to Eq.~(\ref{eq:normal}). The orthonormality condition is numerically confirmed by calculating the non-diagonal elements on a $15\times15$ $\bm q$-mesh for monolayer CrI$_3$, and it is found that maximum absolute value is less than $0.005$. The magnon Berry curvatures can then be estimated using the Wilson-loop approach, with the results being shown in Fig.~\ref{fig:berry}. It is found that the magnon Berry curvatures are distributed mostly around $K$ and $K'$ points and have opposite signs in the two bands. The integral Chern number for the upper and lower magnon bands are $\mathcal{C}=+1$ and $\mathcal{C}=-1$ respectively, demonstrating the non-trivial topology of magnon bands in monolayer CrI$_3$.

\textit{Summary---.} We develop a fully first-principles approach for spin dynamics based on DFPT, from which magnon dispersions as well as magnon wavefunctions can be acquired. We demonstrate that monolayer CrI$_3$ is a magnon Chern insulator where the spin-orbit coupling plays a crucial role. Our model-free approach can also be applied to other topological magnon systems as well as the newly discovered altermagnets~\cite{Smejkal2022}, and thus offers an opportunity to obtain direct microscopic insights into magnons in real materials.

From the viewpoint of methodology, our approach is rigorously exact within the DFPT framework and no prior assumptions are required. Both the adiabatic spin dynamics~\cite{Niu1998,Niu1999,Lin2025} and the Landau-Lifshitz equation can be derived from Eq.~(\ref{eq:eigen1}) with appropriate approximations~\cite{Qian2002}. Moreover, the employment of Sternheimer equation is crucial in our approach, which overcomes the convergence problem with respect to unoccupied bands. As a result, the Goldstone theorem is guaranteed in the absence of spin-orbit coupling and no artificial correction~\cite{Olsen2021} is required. Our approach can be easily implemented in other DFPT codes and can also be generalized to many-body perturbation theory.

\begin{acknowledgments}
\textit{Acknowledgments---.} This work was financially supported by the National Key R\&D Program of China (Grant No. 2024YFA1408103), National Natural Science Foundation of China (12474158, 12234017, 12488101 and 12404288), Innovation Program for Quantum Science and Technology (2021ZD0302800), Anhui Initiative in Quantum Information Technologies (AHY170000), Postdoctoral Fellowship Program of CPSF (No. GZC20232561), China Postdoctoral Science Foundation (2023M733413) and the Fundamental Research Funds for the Central Universities (WK9990000132). We also thank the Supercomputing Center of University of Science and Technology of China for providing high-performance computing resources.
\end{acknowledgments}

% \subsection*{A.1 Cotangent bundle}

%\bibliographystyle{plain}
%\bibliography{magnon}

\end{document}